# $SU(3)_C \times SU(2)_L \times U(1)_Y \, (\times U(1)_X)$ as a symmetry of division algebraic ladder operators

C. Furey[1,2,a]

[1] Department of Applied Mathematics and Theoretical Physics, University of Cambridge, Wilberforce Road, Cambridge CB3 0WA, UK
[2] Cavendish Laboratory, University of Cambridge, JJ Thomson Avenue, Cambridge CB3 0HE, UK

**Abstract** We demonstrate a model which captures certain attractive features of $SU(5)$ theory, while providing a possible escape from proton decay. In this paper we show how ladder operators arise from the division algebras $\mathbb{R}$, $\mathbb{C}$, $\mathbb{H}$, and $\mathbb{O}$. From the $SU(n)$ symmetry of these ladder operators, we then demonstrate a model which has much structural similarity to Georgi and Glashow's $SU(5)$ grand unified theory. However, in this case, the transitions leading to proton decay are expected to be blocked, given that they coincide with presumably forbidden transformations which would incorrectly mix distinct algebraic actions. As a result, we find that we are left with $G_{sm} = SU(3)_C \times SU(2)_L \times U(1)_Y/\mathbb{Z}_6$. Finally, we point out that if $U(n)$ ladder symmetries are used in place of $SU(n)$, it may then be possible to find this same $G_{sm} = SU(3)_C \times SU(2)_L \times U(1)_Y/\mathbb{Z}_6$, together with an extra $U(1)_X$ symmetry, related to $B-L$.

## 1 Introduction

From a wide range of possible theories, the standard model has emerged almost uniquely, after having survived decades of experimental scrutiny. And so this raises the question: What makes $SU(3)_C \times SU(2)_L \times U(1)_Y/\mathbb{Z}_6$ so special? Of the infinite number of imaginable gauge groups, why *this* gauge group?

Furthermore, as we know, a choice of gauge group does not imply a complete description of a theory. Even if we were to understand why nature's local symmetries should be given by $G_{sm} \equiv SU(3)_C \times SU(2)_L \times U(1)_Y/\mathbb{Z}_6$, we would still be at a loss to explain the standard model's particle content. Clearly, the requirement of anomaly cancellation alone cannot go far enough to narrow down the possibilities.

However, despite this embarrassment of riches, the standard model's fermions are nonetheless identified with a specific choice of $SU(3)_C \times SU(2)_L \times U(1)_Y/\mathbb{Z}_6$ representations. These are given by $\mathcal{L} = (1, 2, -1/2)$ for left-handed leptons, $\mathcal{Q} = (3, 2, 1/6)$ for left-handed quarks, $\mathcal{E} = (1, 1, -1)$ for the right-handed electron, $\mathcal{U} = (3, 1, 2/3)$ for right-handed up quarks, and $\mathcal{D} = (3, 1, -1/3)$ for right-handed down quarks. These irreducible representations must then be replicated twice, so as to account for the standard model's three generations.

So in addition to deciphering the reasoning behind the standard model's curious gauge group, it is furthermore upon us to decipher the reasoning behind its curious list of irreducible representations.

Now, in 1974, H. Georgi and S. Glashow introduced one of the first grand unified theories, based on the 24-dimensional gauge group $SU(5)$. The group $SU(5)$ is the smallest simple Lie group to contain $G_{sm}$, admit complex representations, accommodate the standard model's particle content, and be free of anomalies [1,2]. Hence, it offers perhaps the most natural simplification of particle physics at high energies.

More importantly, $SU(5)$ theory does reveal at least a couple of clues about the standard model's mysterious structure. For example, we find that the standard model's peculiar list of hypercharges is explained when $G_{sm}$ is embedded into $SU(5)$ [3,4]. Furthermore, some modern versions of the Georgi–Glashow model have $SU(5)$ acting on the 32-$\mathbb{C}$-dimensional exterior algebra $\Lambda\mathbb{C}^5$ (as opposed to the originally proposed $5^*$ and 10 irreps). The exterior algebra $\Lambda\mathbb{C}^5$ breaks down into the $1 \oplus 5 \oplus 10 \oplus 10^* \oplus 5^* \oplus 1$ irreducible representations of $SU(5)$. This successfully compiles a full generation of quarks and leptons into a single object, $\Lambda\mathbb{C}^5$, while accounting for their anti-particles, and including a right-handed neutrino.[1]

However, with this being said, we note that $SU(5)$ theory does not truly explain the origin of $G_{sm}$. In the typi-cal scenario, $SU(5)$ breaks down to $G_{sm}$ via spontaneous symmetry breaking, mediated by a cleverly chosen Higgs field and potential. The question: *Why $SU(3)_C \times SU(2)_L \times U(1)_Y/\mathbb{Z}_6$?* in this theory is then merely replaced by *Why $SU(5)$?* and *Why this Higgs?*

---

[1] It should be noted that accounting for particles and anti-particles separately is generally thought of as double counting.

[a] e-mail: nf252@cam.ac.uk

To make matters worse, the extra generators of $SU(5)$ enable transitions which cause the proton to decay. Calculations of the proton lifetime within minimal $SU(5)$ theory vary depending on source [5], but are generally considered to be at odds with experiment. This conflict was confirmed recently by the Super-Kamiokande collaboration [6,7], which thus far has turned up no evidence in support of the proton's decay.

As a result, minimal $SU(5)$ theory is largely believed to be ruled out. However, given its strengths, one might be led to wonder if perhaps there could be some missing mathematical structure which ultimately saves $SU(5)$ theory from itself.

In this paper, we begin with four special algebras: the real numbers, $\mathbb{R}$, the complex numbers, $\mathbb{C}$, the quaternions, $\mathbb{H}$, and the octonions, $\mathbb{O}$. Uniquely, these are identified as the only four *normed division algebras over the real numbers*. They are of dimensions 1, 2, 4, and 8 respectively.

Using only $\mathbb{R}$, $\mathbb{C}$, $\mathbb{H}$, and $\mathbb{O}$, we construct a faithful representation of the Clifford algebra $\mathbb{C}l(10)$, acting on a 32-$\mathbb{C}$-dimensional spinor. This 32-$\mathbb{C}$-dimensional spinor is built as a minimal left ideal, which, for our purposes, can be seen to be equivalent to $\Lambda\mathbb{C}^5$. We then propose to identify the model's gauge symmetry with the special unitary symmetry of $\mathbb{C}l(10)$ ladder operators. For $\mathbb{C}l(10)$, these ladder symmetries are identified as $SU(5)$. Consequently, we obtain a division algebraic representation of Georgi and Glashow's $SU(5)$ model.

However, upon closer inspection, we argue that $SU(5)$ symmetry should never be fully realised in this division algebraic construction. Instead, the new underlying algebraic structure is seen to block certain transitions *under the assumption that conceptually distinct algebraic actions do not mix*. Incidentally, these are the transitions responsible for proton decay. In place of $SU(5)$, we are then left with a symmetry group given by $SU(3)_C \times SU(2)_L \times U(1)_Y/\mathbb{Z}_6$.

This work builds on an early finding [8], that the octonions break down into $1 \oplus 3 \oplus 3^* \oplus 1$ irreducible representations under $SU(3) \subset Aut(\mathbb{O}) = G_2$. In their paper, Günaydin and Gürsey identified the 3 and the $3^*$ as a triplet of quarks and anti-quarks under the colour group $SU(3)_C$.

Since that time, a number of authors have expanded on these early results, notably [9–11]. The authors [9,10] were able to find significant pieces of standard model structure by considering carefully chosen tensor products of Clifford algebras, whereas [11] identified significant pieces of standard model structure starting from the algebra $\mathbb{R} \otimes \mathbb{C} \otimes \mathbb{H} \otimes \mathbb{O}$ augmented to $2 \times 2$ matrices. Readers are encouraged to consult the work of these earlier authors.

This paper contributes to the existing literature by exposing a rather straightforward path from $\mathbb{R}$, $\mathbb{C}$, $\mathbb{H}$, and $\mathbb{O}$ to the standard model's gauge group, $SU(3)_C \times SU(2)_L \times U(1)_Y/\mathbb{Z}_6$, with the possibility of an extra $U(1)_X$ symmetry, related to $B-L$. Furthermore, we find that the stable subspaces (minimal ideals) of this division algebraic model exhibit the behaviour of one full generation of quarks and leptons, supplemented with a right-handed neutrino.

It is hoped that these results will be of direct use to those currently working on grand unified theories [12–19], extra dimensions [20–24], and non-commutative geometry [25–29]. Furthermore, it may lend helpful clues to those working in related fields such as supersymmetry [30–36], and other topics closely related to the division algebras [37–40]. Finally, the possibility of an extra $B-L$ symmetry mentioned here is particularly exciting given the recent findings of [41–44].

## 2 The Georgi–Glashow model

### 2.1 $SU(5)$ acting on $\Lambda\mathbb{C}^5$

The $SU(5)$ model is perhaps the most logical choice for a grand unified theory. With its 24 dimensions, $SU(5)$ is the smallest acceptable Lie group in which the 12-dimensional $G_{sm}$ can be embedded. From $SU(5)$'s list of irreducible representations, Georgi and Glashow selected the $5^*$ and 10 so as to portray the standard model's 15 quarks and leptons.

Having said that, more modern versions of the theory have since included the singlet of $SU(5)$, which plays the part of a sterile right-handed neutrino. These representations, $1 \oplus 5^* \oplus 10$, combine naturally into the exterior algebra $\Lambda\mathbb{C}^5 \sim 1 \oplus 5^* \oplus 10 \oplus 10^* \oplus 5 \oplus 1$. Clearly, $\Lambda\mathbb{C}^5$ now also includes representations corresponding to the anti-particles of the $1 \oplus 5^* \oplus 10$ states. In Fig. 1, we depict the irreducible representations of $SU(5)$ within $\Lambda\mathbb{C}^5$.

Parenthetically, we mention that $\Lambda\mathbb{C}^5$ can alternately split into two 16-$\mathbb{C}$-dimensional irreducible representations under $Spin(10)$, for the "$SO(10)$" grand unified theory. In this case, one irreducible representation corresponds to the even-graded subspace of $\Lambda\mathbb{C}^5$, and the other corresponds to its odd-graded subspace. Furthermore, we point out that the full 32-complex-dimensional $\Lambda\mathbb{C}^5$ provides the only irreducible representation for the complex Clifford algebra, $\mathbb{C}l(10)$.

Already at the level of $SU(5)$ acting on $\Lambda\mathbb{C}^5$, one might begin to suspect that the standard model's particle content is not entirely arbitrary. It seems too much a coincidence to think that $\Lambda\mathbb{C}^5$ should gratuitously provide the perfect space for one full generation, and that $SU(5)$ should naturally include $SU(3)_C \times SU(2)_L \times U(1)_Y/\mathbb{Z}_6$.

| | | | | | |
|---|---|---|---|---|---|
| | | $\alpha_1\alpha_2\alpha_3\alpha_4\alpha_5$ | | | **1** |
| | $\alpha_1\alpha_2\alpha_3\alpha_4$ | ... | $\alpha_2\alpha_3\alpha_4\alpha_5$ | | **5**$^*$ |
| $\alpha_1\alpha_2\alpha_3$ | $\alpha_1\alpha_2\alpha_4$ | ... | $\alpha_2\alpha_4\alpha_5$ | $\alpha_3\alpha_4\alpha_5$ | **10**$^*$ |
| $\alpha_1\alpha_2$ | $\alpha_1\alpha_3$ | ... | $\alpha_3\alpha_5$ | $\alpha_4\alpha_5$ | **10** |
| $\alpha_1$ | $\alpha_2$ | $\alpha_3$ | $\alpha_4$ | $\alpha_5$ | **5** |
| | | 1 | | | **1** |

**Fig. 1** The exterior algebra $\Lambda\mathbb{C}^5$, broken down in terms of $SU(5)$ irreducible representations. Here, the basis elements within a given irreducible representation appear within a single row. Starting from the bottom and moving upward, the grade-0 object, 1, forms an $SU(5)$ singlet, the grade-1 objects, $\alpha_i$, form the 5, the grade-2 objects, $\alpha_i\alpha_j$, form the 10, the grade-3 objects, $\alpha_i\alpha_j\alpha_k$, form the 10*, the grade-4 objects, $\alpha_i\alpha_j\alpha_k\alpha_\ell$, form the 5*, and the grade-5 object, $\alpha_1\alpha_2\alpha_3\alpha_4\alpha_5$, forms another singlet

### 2.2 Breaking $SU(5)$

Now with this being said, $SU(5)$ does indeed condone proton decay, a process thought never to have been observed. For this reason, it is standard practice to then propose an additional Higgs field so as to break $SU(5) \mapsto G_{sm}$ at high energy.

The typical representation chosen for this task is the 24-dimensional adjoint Higgs [5]. Its vacuum expectation value is then made to be proportional to the generator of $U(1)_Y$.

For gauge bosons, spontaneous symmetry breaking induces

$$\underline{24} \mapsto (\underline{8}, \underline{1}, 0) \oplus (\underline{1}, \underline{3}, 0) \oplus (\underline{1}, \underline{1}, 0)$$
$$\oplus \left(\underline{3}, \underline{2}^*, -\frac{5}{6}\right) \oplus \left(\underline{3}^*, \underline{2}, \frac{5}{6}\right), \quad (1)$$

where the $(\underline{8}, \underline{1}, 0)$, $(\underline{1}, \underline{3}, 0)$, and $(\underline{1}, \underline{1}, 0)$ generate $SU(3)_C$, $SU(2)_L$, and $U(1)_Y$, respectively. The $\left(\underline{3}, \underline{2}^*, -\frac{5}{6}\right)$ and $\left(\underline{3}^*, \underline{2}, \frac{5}{6}\right)$ generators give rise to 12 additional gauge bosons which can be seen to mediate proton decay.

As for fermions, the singlet remains unchanged, while the 5* and the 10 break as

$$\underline{5}^* \mapsto \underbrace{\left(\underline{3}^*, \underline{1}, \tfrac{1}{3}\right)}_{\bar{d}_L} \oplus \underbrace{\left(\underline{1}, \underline{2}^*, -\tfrac{1}{2}\right)}_{\ell_L}, \quad (2)$$

$$\underline{10} \mapsto \underbrace{\left(\underline{3}^*, \underline{1}, -\tfrac{2}{3}\right)}_{\bar{u}_L} \oplus \underbrace{\left(\underline{3}, \underline{2}, \tfrac{1}{6}\right)}_{q_L} \oplus \underbrace{\left(\underline{1}, \underline{1}, 1\right)}_{e_L^+}.$$

Finally, the familiar Higgs field $\phi$, responsible for breaking $G_{sm} \mapsto SU(3)_C \times U(1)_{em}$, is often embedded in the 5 of $SU(5)$. At the GUT scale, this breaks as

$$\underline{5} \mapsto \underbrace{\left(\underline{3}, \underline{1}, -\tfrac{1}{3}\right)}_{\mathcal{H}} \oplus \underbrace{\left(\underline{1}, \underline{2}, \tfrac{1}{2}\right)}_{\phi}, \quad (3)$$

where the $\left(\underline{3}, \underline{1}, -\tfrac{1}{3}\right)$ describes a new triplet Higgs field, $\mathcal{H}$.

### 2.3 $SU(5)$ summary

In searching for a minimalistic model of particle physics, one would be hard-pressed to surpass $SU(5)$ acting on $\Lambda\mathbb{C}^5$. Its particle content is concisely defined, and its simple gauge group comes as close as could be expected to that of the standard model of particle physics.

However, compliance with experiment prompts the introduction of a carefully chosen 24-dimensional Higgs field and potential. This breaks $SU(5) \mapsto G_{sm}$, but in the process, also compromises the simplicity of the original model.

More troublesome still, we find that $SU(5)$ theory conflicts with experiment, even with this adaptation. For example, experimental lack of proton decay, lack of 't Hooft-Polyakov monopoles, the doublet-triplet splitting problem, and inaccurate mass predictions have all taken their toll on $SU(5)$ theory to varying degrees [5]. However, given the virtues of $SU(5)$ theory, one might wonder if it could be possible to construct a similar model which can bypass these hazards.

We will now build up a division algebraic representation of the Georgi–Glashow model, piece by piece. Along the way, we will encounter several group representations familiar from particle theory. We begin with the complex quaternions, $\mathbb{C} \otimes \mathbb{H}$.

## 3 $\mathbb{C} \otimes \mathbb{H}$: spin and chirality

### 3.1 Introduction to $\mathbb{C} \otimes \mathbb{H}$

A generic element of $\mathbb{C} \otimes \mathbb{H}$ is written $c_0\epsilon_0 + c_1\epsilon_1 + c_2\epsilon_2 + c_3\epsilon_3$, where the $c_i \in \mathbb{C}$ and $\epsilon_0 \equiv 1$. The basis vectors $\epsilon_1, \epsilon_2, \epsilon_3$ follow the associative, but non-commutative quaternionic multiplication rules

$$\epsilon_1\epsilon_1 = \epsilon_2\epsilon_2 = \epsilon_3\epsilon_3 = \epsilon_1\epsilon_2\epsilon_3 = -1, \quad (4)$$

from which we obtain $\epsilon_1\epsilon_2 = -\epsilon_2\epsilon_1 = \epsilon_3$, $\epsilon_2\epsilon_3 = -\epsilon_3\epsilon_2 = \epsilon_1$, $\epsilon_3\epsilon_1 = -\epsilon_1\epsilon_3 = \epsilon_2$.

We define two notions of conjugation on an element $a \in \mathbb{C} \otimes \mathbb{H}$. The *complex conjugate* of $a$, denoted $a^*$, maps the complex $i$ to $-i$. That which we will call the *hermitian conjugate* of $a$, denoted $a^\dagger$, maps $i$ to $-i$ and $\epsilon_j$ to $-\epsilon_j$ for $j = 1, 2, 3$, while reversing the order of multiplication, $(ab)^\dagger = b^\dagger a^\dagger$.

A well-known correspondence exists between the complex quarternions and the Pauli matrices. That is,

$$i\epsilon_1 \leftrightarrow \sigma_x, \quad i\epsilon_2 \leftrightarrow \sigma_y, \quad i\epsilon_3 \leftrightarrow \sigma_z. \quad (5)$$

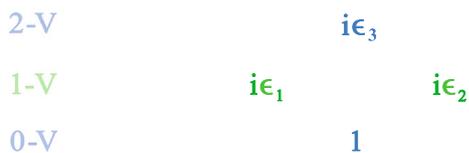

**Fig. 2** The 4-complex-dimensional $\mathbb{C} \otimes \mathbb{H}$ gives a faithful representation of $\mathbb{C}l(2)$. The pair of vectors, $i\epsilon_1$ and $i\epsilon_2$, form the generating space (1-vectors) of the Clifford algebra, and $\epsilon_3 = \epsilon_1 \epsilon_2$ is a bi-vector

However, readers should note that these objects behave more symmetrically than do the Pauli matrices under complex conjugation. Explicitly, $(i\epsilon_j)^* = -i\epsilon_j \; \forall j$ versus $\sigma_x^* = +\sigma_x$, $\sigma_y^* = -\sigma_y$, $\sigma_z^* = +\sigma_z$.

### 3.2 Clifford algebraic structure

It is straightforward to confirm that the left action of $\mathbb{C} \otimes \mathbb{H}$ on itself gives a faithful representation of the complex Clifford algebra $\mathbb{C}l(2)$. Please see Fig. 2. Here the generating vectors $i\epsilon_1$ and $i\epsilon_2$ obey $\{i\epsilon_m, i\epsilon_n\} = 2\delta_{mn}$ for $m = 1, 2$ and $n = 1, 2$.

Now, the generating space given in Fig. 2 may be rewritten in terms of a new basis,

$$\alpha \equiv \frac{1}{2}(\epsilon_2 + i\epsilon_1), \quad \alpha^\dagger \equiv \frac{1}{2}(-\epsilon_2 + i\epsilon_1). \tag{6}$$

Under anti-commutation, these operators behave as

$$\{\alpha, \alpha\} = \{\alpha^\dagger, \alpha^\dagger\} = 0, \quad \{\alpha, \alpha^\dagger\} = 1. \tag{7}$$

Since $\alpha$ and $\alpha^\dagger$ span the Clifford algebra's generating space, we then see that all of $\mathbb{C} \otimes \mathbb{H} \simeq \mathbb{C}l(2)$ may be described as sums and multiples of these ladder operators.

### 3.3 Right-handed Weyl spinors as minimal left ideals

The Clifford algebraic structure of $\mathbb{C} \otimes \mathbb{H}$ is important for us since it will allow us to construct Weyl spinors. That is, we will make use of the fact that spinors can be defined as *minimal left ideals* of Clifford algebras [45].

Given an algebra, $A$, a *left ideal*, $B$, is a subalgebra of $A$ whereby $ab$ is in $B$ for all $b$ in $B$, and for any $a$ in $A$. Said another way, the subspace $B$ is stable under left multiplication by any of the elements $a \in A$. Now, a *minimal* left ideal is a left ideal which contains no left ideals other than $\{0\}$ and itself. (This choice of definition for spinors was motivated by the algebraic path integral model described in Chapter 2 of [46], where particles are identified as surviving subspaces of some fundamental algebra.)

We will now construct spinors as minimal left ideals, largely following the procedure set out in [45] for Clifford algebras $\mathbb{C}l(2n)$ with $n \in \mathbb{Z} > 0$. In this construction, the first task is to build a particular idempotent, $v = vv$. Then the spinor is obtained by simply left multiplying $\mathbb{C}l(2n)$ onto $v$, as in $\Psi \equiv \mathbb{C}l(2n)v$.

In our current case of $\mathbb{C}l(2)$, our idempotent will be defined as $v_s \equiv \alpha\alpha^\dagger$, where the subscript $s$ refers to spin. For generic $\mathbb{C}l(2n)$, the Clifford algebra's generating space will be split into $n$ operators $\alpha_i$ and $n$ operators $\alpha_i^\dagger$. In this more general case, the idempotent $v$ will be constructed as $v \equiv \alpha_1 \alpha_2 \ldots \alpha_n \alpha_n^\dagger \ldots \alpha_2^\dagger \alpha_1^\dagger$.

For $\mathbb{C} \otimes \mathbb{H} \simeq \mathbb{C}l(2)$, readers may confirm that the resulting minimal left ideal $\mathbb{C}l(2)v_s$ takes the form

$$\Psi_R \equiv \psi_R^\uparrow \alpha^\dagger v_s + \psi_R^\downarrow v_s, \tag{8}$$

where $\psi_R^\uparrow$ and $\psi_R^\downarrow$ are complex coefficients. Here, we have labelled the basis vector $\alpha^\dagger v_s$ as spin-up since $\frac{1}{2}i\epsilon_3\alpha^\dagger v_s = \frac{1}{2}\alpha^\dagger v_s$, and we have labelled $v_s$ as spin-down since $\frac{1}{2}i\epsilon_3 v_s = -\frac{1}{2}v_s$. We have furthermore labeled the spinor $\Psi_R$ as right-handed, which may be viewed as an arbitrary choice at this point.

Readers may notice the resemblance between $\Psi_R$ and a Fock space, where $v_s$ formally plays the role of a vacuum state.

### 3.4 Left-handed Weyl spinors as minimal left ideals

Incidentally, another spinor may be constructed in $\mathbb{C}l(2)$ by swapping the roles of $\alpha$ and $\alpha^\dagger$. In this case, let us then define $v'_s \equiv \alpha^\dagger \alpha$. When expressed in terms of $\mathbb{C} \otimes \mathbb{H}$, it so happens that $v'_s$ is given by $v'_s = v_s^*$. From here, we can construct a minimal left ideal, linearly independent from the first.

$$\Psi_L \equiv \psi_L^\uparrow v_s^* + \psi_L^\downarrow \alpha v_s^*, \tag{9}$$

where $\psi_L^\uparrow$ and $\psi_L^\downarrow$ are complex coefficients.

The identification of $\Psi_L$ and $\Psi_R$ as left- and right-handed Weyl spinors is justified when we take $\gamma_5$ to be represented as right multiplication by $-i\epsilon_3$ as shown in Section 4.7 of [46]. Furthermore, in Chapter 3 of [46], $\Psi_L$ is shown to transform as does a left-handed Weyl spinor under $SL(2, \mathbb{C})$, $\Psi'_L = L\Psi_L$, and $\Psi_R$ is shown to transform as does a right-handed Weyl spinor, $\Psi'_R = L^* \Psi_R$. Here, $L \in \mathbb{C} \otimes \mathbb{H}$ is defined as $L \equiv \exp(r_j \epsilon_j + b_j i \epsilon_j)$, where $j = 1, 2, 3$, and $r_j, b_j \in \mathbb{R}$.

Putting both subspaces together, we may define Dirac spinors as $\Psi_D \equiv \Psi_L + \Psi_R$. When translated into the formalism of $2 \times 2$ $\mathbb{C}$ matrices via relations (5), this gives

$$\Psi_D \mapsto \begin{pmatrix} \psi_L^\uparrow & \psi_R^\uparrow \\ \psi_L^\downarrow & \psi_R^\downarrow \end{pmatrix}, \tag{10}$$

so that each Weyl spinor occupies a column within the $2 \times 2$ $\mathbb{C}$ matrices. Multiplying from the left induces rotations between spin states, while multiplication from the right induces rotation between chiralities.

In this paper, we will sometimes alternate between matrix descriptions and division algebraic descriptions of our states. Having said that, readers should take note that not all descriptions are created equal. After all, it is the division algebras, not the matrix algebras, which ultimately dictate which Clifford algebras we will consider. Also, the familiar operation of charge conjugation finds a more succinct description within this division algebraic formalism, as we will now show.

### 3.5 Complex conjugation and charge conjugation

To give a basis for comparison, let us first consider the Dirac spinor, as described by $2 \times 2$ $\mathbb{C}$ matrices in relation (10). Under the action of complex conjugation, $i \mapsto -i$, we simply conjugate the four complex coefficients, a procedure which does not lead to anything of particular significance.

In contrast, let us now apply the map $i \mapsto -i$ to $\Psi_D$ when it is described in terms of $\mathbb{C} \otimes \mathbb{H}$,

$$\begin{aligned}\Psi_D^* &= \Psi_L^* + \Psi_R^* \\ &= \left(\psi_L^\uparrow v_s^* + \psi_L^\downarrow \alpha v_s^*\right)^* + \left(\psi_R^\uparrow \alpha^\dagger v_s + \psi_R^\downarrow v_s\right)^* \\ &= \left(\psi_R^{\downarrow *} v_s^* - \psi_R^{\uparrow *}\alpha v_s^*\right) + \left(-\psi_L^{\downarrow *}\alpha^\dagger v_s + \psi_L^{\uparrow *} v_s\right).\end{aligned} \quad (11)$$

This transformation may be recognizable to readers as $\psi \mapsto \psi_c = -i\gamma_2 \psi^*$ from the standard formalism of quantum field theory (up to a phase). *In other words, $i \mapsto -i$ in the $\mathbb{C} \otimes \mathbb{H}$ formalism yields charge conjugation on Dirac spinors.*

In this light, we may see that the role of $i\gamma_2$ in the standard formalism is to take into account the complex conjugation of basis vectors in the $\mathbb{C} \otimes \mathbb{H}$ formalism. When spinors are written in terms of $\mathbb{C} \otimes \mathbb{H}$, the object $i\gamma_2$ need no longer be put in by hand.

### 3.6 Dirac algebra

The Weyl spinors $\Psi_L$ and $\Psi_R$ were each constructed as minimal ideals of $\mathbb{C} \otimes \mathbb{H}$, under *left* multiplication. However, transitions *between* $\Psi_L$ and $\Psi_R$ can be effected via right multiplication. This right action provides another faithful representation of $\mathbb{C}l(2)$.

$$\begin{array}{ccc}\mathbb{C}\otimes\mathbb{H} & \triangleright \;\; \Psi_D \;\; \triangleleft & \mathbb{C}\otimes\mathbb{H} \\ \downarrow & & \downarrow \\ \mathbb{C}l(2) & & \mathbb{C}l(2) \\ \text{spin} & & \text{chirality} \end{array} \quad (12)$$

Hence the combined action of both left and right multiplication gives $\mathbb{C}l(2) \otimes_{\mathbb{C}} \mathbb{C}l(2) \simeq \mathbb{C}l(4)$. From these actions, we may construct the $\mathbb{C} \otimes \mathbb{H}$-equivalent of the Dirac matrices. Generators of the Dirac algebra in the Weyl basis may be described as

$$\begin{aligned}&\gamma^0 = 1|i\epsilon_1 \quad \gamma^1 = i\epsilon_1|\epsilon_2 \\ &\gamma^2 = i\epsilon_2|\epsilon_2 \quad \gamma^3 = i\epsilon_3|\epsilon_2,\end{aligned} \quad (13)$$

as introduced in Section 4.7 of [46]. Here, we made use of the bar notation of [47]. By definition, the operator $x|y$ acting on some element $z$, for $x, y, z \in \mathbb{C} \otimes \mathbb{H}$, is given by $xzy$.

We end this section by pointing out that $\mathbb{C} \otimes \mathbb{H}$ is capable of describing more than just Weyl and Dirac spinors. This 4-$\mathbb{C}$-dimensional algebra was shown in [46] to also describe Majorana spinors, scalars, four-vectors, and the field strength tensor - each in the form of generalized ideals. By generalized ideals, we mean *invariant subspaces under some action of the algebra on itself*. These account for all of the Lorentz representations of the standard model.

For a recent electromagnetic model which builds on this formalism, see [48].

### 3.7 $\mathbb{C} \otimes \mathbb{H}$ summary

In this section, we showed that the left action of $\mathbb{C} \otimes \mathbb{H}$ on itself gives a faithful representation of the Clifford algebra $\mathbb{C}l(2)$. We then identified anti-commuting ladder operators generating $\mathbb{C}l(2)$, and used them to build a pair of minimal left ideals, $\Psi_L$ and $\Psi_R$. These two Weyl spinors may then be combined as $\Psi_D = \Psi_L + \Psi_R$ so as to give a single irreducible representation when both the left and right actions of $\mathbb{C} \otimes \mathbb{H}$ are considered. This results in a faithful representation of $\mathbb{C}l(2) \otimes_{\mathbb{C}} \mathbb{C}l(2) \simeq \mathbb{C}l(4) \simeq \mathbb{C} \otimes Cl(1, 3)$. In short, $\mathbb{C} \otimes \mathbb{H}$ lends itself naturally to the description of those spinors familiar to $3 + 1$ spacetime dimensions.

## 4 $\mathbb{C} \otimes \mathbb{O}$: colour

We will now repeat this construction for the case of the complex octonions. In analogy to the $\mathbb{C} \otimes \mathbb{H}$ minimal left ideals $\Psi_L + \Psi_R$, we will construct $\mathbb{C} \otimes \mathbb{O}$ minimal left ideals, $S^u + S^d$. Subsequently, we will find that $S^u$ and $S^d$ mirror the behaviour of one generation of quarks and leptons under $SU(3)_C$.

### 4.1 Introduction to $\mathbb{C} \otimes \mathbb{O}$

A generic element of $\mathbb{C} \otimes \mathbb{O}$ is written $\sum_{n=0}^{7} c_n e_n$, where the $c_n \in \mathbb{C}$. The $e_n$ are octonionic imaginary units $(e_n^2 = -1)$, apart from $e_0 \equiv 1$. The multiplication rules for these imaginary units can be defined by setting $e_1 e_2 = e_4$, and then applying the following rules,

$$e_i e_j = -e_j e_i \quad i \neq j,$$
$$e_i e_j = e_k \Rightarrow e_{i+1} e_{j+1} = e_{k+1},$$
$$e_i e_j = e_k \Rightarrow e_{2i} e_{2j} = e_{2k}. \tag{14}$$

Alternately, readers may consult [46,49], or [50], where the given Fano plane depicts these same multiplication rules. The octonions form a non-associative algebra, meaning that the relation $(ab)c = a(bc)$ does not always hold. Octonionic automorphisms are given by $G_2$, the 14-dimensional exceptional Lie group.

### 4.2 Clifford algebraic structure

In parallel with the case of $\mathbb{C} \otimes \mathbb{H}$, we will now consider the left action of $\mathbb{C} \otimes \mathbb{O}$ on itself. It can be confirmed that complex linear combinations of octonions repeatedly left multiplying $f \in \mathbb{C} \otimes \mathbb{O}$ may always be written in the canonical form

$$Mf \equiv c_0 f + \sum_{i=1}^{6} c_i e_i f + \sum_{j=2}^{6} \sum_{i=1}^{j-1} c_{ij} e_i(e_j f)$$
$$+ \sum_{k=3}^{6} \sum_{j=2}^{k-1} \sum_{i=1}^{j-1} c_{ijk} e_i(e_j(e_k f)) + \cdots$$
$$+ c_{123456} e_1(e_2(e_3(e_4(e_5(e_6 f))))), \tag{15}$$

where the coefficients $c_0, c_i, \ldots \in \mathbb{C}$. Readers may note that the octonionic imaginary unit $e_7$ is not explicitly expressed in these maps. This is due to the fact that

$$e_7 f = e_1(e_2(e_3(e_4(e_5(e_6 f))))) \quad \forall f \in \mathbb{C} \otimes \mathbb{O}, \tag{16}$$

thereby making $e_7$ redundant as a left-action map. Of course, $e_7$ itself holds no preferred status within the octonions, and the space of left-action maps may equivalently be described by chains built from any six of the seven imaginary units. By using the identity (16), the Eq. (15) may then be written more compactly as

$$Mf = c_0 f + \sum_{i=1}^{7} c_i \overleftarrow{e_i} f + \sum_{j=2}^{7} \sum_{i=1}^{j-1} c_{ij} \overleftarrow{e_i e_j} f$$
$$+ \sum_{k=3}^{7} \sum_{j=2}^{k-1} \sum_{i=1}^{j-1} c_{ijk} \overleftarrow{e_i e_j e_k} f, \tag{17}$$

where the arrows indicate the direction of bracketing, as in $\overleftarrow{e_i e_j e_k} f = \overleftarrow{e_i} \overleftarrow{e_j} \overleftarrow{e_k} f = e_i(e_j(e_k f))$.

It can be verfied that

$$\overleftarrow{e_i} \overleftarrow{e_j} f + \overleftarrow{e_j} \overleftarrow{e_i} f = e_i(e_j f) + e_j(e_i f) = -2\delta_{ij} f \tag{18}$$

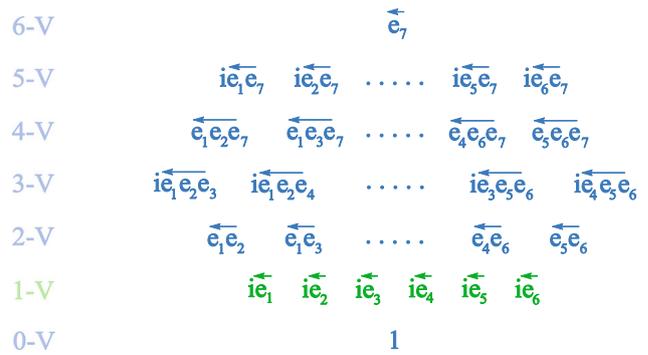

**Fig. 3** The left action of $\mathbb{C} \otimes \mathbb{O}$ on itself provides a faithful representation of $\mathbb{C}l(6)$, and is hence isomorphic to the $8 \times 8$ $\mathbb{C}$ matrices. Here, left-action maps, $i\overleftarrow{e_j}$, for $j = 1, \ldots 6$, form the generating space (1-vectors)

for $i, j = 1, \ldots 6$, and $\forall f \in \mathbb{C} \otimes \mathbb{O}$, and furthermore that the left action of $\mathbb{C} \otimes \mathbb{O}$ on itself gives a faithful representation of the Clifford algebra $\mathbb{C}l(6)$. Please see Fig. 3. Here, multiplication is understood to be given by the composition of left action maps, and hence is associative by definition. The Clifford algebra $\mathbb{C}l(6)$ is isomorphic to the $8 \times 8$ $\mathbb{C}$ matrices. This scenario closely emulates our earlier example where the left action of $\mathbb{C} \otimes \mathbb{H}$ on itself gave a faithful representation of the Clifford algebra $\mathbb{C}l(2)$, isomorphic to the $2 \times 2$ $\mathbb{C}$ matrices.

Now, the generating space given in Fig. 3 may be rewritten in terms of a new basis,

$$\overleftarrow{a_1} \equiv \frac{1}{2}\left(-\overleftarrow{e_5} + i\overleftarrow{e_4}\right), \quad \overleftarrow{a_2} \equiv \frac{1}{2}\left(-\overleftarrow{e_3} + i\overleftarrow{e_1}\right),$$
$$\overleftarrow{a_3} \equiv \frac{1}{2}\left(-\overleftarrow{e_6} + i\overleftarrow{e_2}\right), \quad \overleftarrow{a_1}^\dagger \equiv \frac{1}{2}\left(\overleftarrow{e_5} + i\overleftarrow{e_4}\right),$$
$$\overleftarrow{a_2}^\dagger \equiv \frac{1}{2}\left(\overleftarrow{e_3} + i\overleftarrow{e_1}\right), \quad \overleftarrow{a_3}^\dagger \equiv \frac{1}{2}\left(\overleftarrow{e_6} + i\overleftarrow{e_2}\right). \tag{19}$$

Here, we define the conjugation † to map $i \mapsto -i$ and $\overleftarrow{e_j} \mapsto -\overleftarrow{e_j}$ for $j = 1 \ldots 7$. As with the hermitian conjugation of matrices, † also reverses the order of multiplication, that is, the order of left action maps: $(\overleftarrow{x}\,\overleftarrow{y})^\dagger = \overleftarrow{y}^\dagger \overleftarrow{x}^\dagger$.

Under anti-commutation, these operators behave as

$$\{\overleftarrow{a_i}, \overleftarrow{a_j}\} f \equiv a_i(a_j f) + a_j(a_i f) = 0,$$
$$\{\overleftarrow{a_i}^\dagger, \overleftarrow{a_j}^\dagger\} f \equiv a_i^\dagger(a_j^\dagger f) + a_j^\dagger(a_i^\dagger f) = 0,$$
$$\{\overleftarrow{a_i}, \overleftarrow{a_j}^\dagger\} f \equiv a_i(a_j^\dagger f) + a_j^\dagger(a_i f) = \delta_{ij} f, \quad \forall f \in \mathbb{C} \otimes \mathbb{O}, \tag{20}$$

which is simply a higher-dimensional analogue of equations (7).

From this point forward, we will not be interested in the object $f \in \mathbb{C} \otimes \mathbb{O}$, only the maps $\overleftarrow{a_i}$ and $\overleftarrow{a_j}^\dagger$ which act on it. Hence, we will no longer refer to $f$ explicitly. Furthermore, in the interest of simplifying notation, we will forfeit the use of arrows on our left action maps, although their presence should be implicitly understood. Equations (20) can then be rewritten more succinctly as

$$\{a_i, a_j\} = 0, \quad \{a_i^\dagger, a_j^\dagger\} = 0, \quad \{a_i, a_j^\dagger\} = \delta_{ij}. \quad (21)$$

### 4.3 Quarks and leptons as minimal left ideals

Now that we have established Clifford algebraic structure generated by $\mathbb{C} \otimes \mathbb{O}$, we would like to build minimal left ideals analogous to $\Psi_L$ and $\Psi_R$. Again following [45], let us define $\omega \equiv a_1 a_2 a_3$ so that our idempotent is then given by $v_c \equiv \omega \omega^\dagger$. Our first minimal left ideal will be given by $S^u \equiv \mathbb{C}l(6)v_c$,

$$\begin{aligned} S^u &= \mathcal{V}v_c \\ &+ \bar{\mathcal{D}}^r a_1^\dagger v_c + \bar{\mathcal{D}}^g a_2^\dagger v_c + \bar{\mathcal{D}}^b a_3^\dagger v_c \\ &+ \mathcal{U}^r a_3^\dagger a_2^\dagger v_c + \mathcal{U}^g a_1^\dagger a_3^\dagger v_c + \mathcal{U}^b a_2^\dagger a_1^\dagger v_c \\ &+ \mathcal{E}^+ a_3^\dagger a_2^\dagger a_1^\dagger v_c, \end{aligned} \quad (22)$$

where $\mathcal{V}, \bar{\mathcal{D}}^r, \ldots \mathcal{E}^+$ are 8 suggestively named complex coefficients.

Swapping the roles of $a_i$ and $a_i^\dagger$ gives a linearly independent ideal, $S^d \equiv \mathbb{C}l(6)\omega^\dagger \omega = \mathbb{C}l(6)v_c^*$,

$$\begin{aligned} S^d &= \bar{\mathcal{V}}v_c^* \\ &- \mathcal{D}^r a_1 v_c^* - \mathcal{D}^g a_2 v_c^* - \mathcal{D}^b a_3 v_c^* \\ &+ \bar{\mathcal{U}}^r a_3 a_2 v_c^* + \bar{\mathcal{U}}^g a_1 a_3 v_c^* + \bar{\mathcal{U}}^b a_2 a_1 v_c^* \\ &+ \mathcal{E}^- a_1 a_2 a_3 v_c^*, \end{aligned} \quad (23)$$

where $\bar{\mathcal{V}}, \mathcal{D}^r, \ldots \mathcal{E}^-$ are eight complex coefficients. The one-generation labeling we have used in $S^u$ and $S^d$ will be partially justified now, and fully justified by the end of this article.

### 4.4 $SU(3)$ ladder symmetry

Let us now consider an $SU(3)$ symmetry acting on the ladder operators, $a_1, a_2, a_3$. Taking $r_j \in \mathbb{R}$, our raising and lowering operators transform as

$$e^{ir_j \Lambda_j} a_k e^{-ir_j \Lambda_j} \quad \text{and} \quad e^{ir_j \Lambda_j} a_k^\dagger e^{-ir_j \Lambda_j}, \quad (24)$$

where the eight $\Lambda_j$ span $su(3)$, and are given by

$$\begin{aligned} &\Lambda_1 = -a_2^\dagger a_1 - a_1^\dagger a_2, \quad \Lambda_2 = i a_2^\dagger a_1 - i a_1^\dagger a_2, \\ &\Lambda_3 = a_2^\dagger a_2 - a_1^\dagger a_1, \quad \Lambda_4 = -a_1^\dagger a_3 - a_3^\dagger a_1, \\ &\Lambda_5 = -i a_1^\dagger a_3 + i a_3^\dagger a_1, \quad \Lambda_6 = -a_3^\dagger a_2 - a_2^\dagger a_3, \\ &\Lambda_7 = i a_3^\dagger a_2 - i a_2^\dagger a_3, \quad \Lambda_8 = -\frac{1}{\sqrt{3}}\left(a_1^\dagger a_1 + a_2^\dagger a_2 - 2a_3^\dagger a_3\right). \end{aligned} \quad (25)$$

This representation of $SU(3)$ is given by the subgroup of $G_2$ which holds the octonionic $e_7$ constant.

Given that our minimal left ideals are built entirely out of ladder operators, we see that transformations on $a_i$ and $a_j^\dagger$ thereby induce transformations on $S^u$ and $S^d$. Under $SU(3)$, $S^u$ and $S^d$ are found to transform as

$$S^u \sim 1 \oplus 3^* \oplus 3 \oplus 1, \quad S^d \sim 1 \oplus 3 \oplus 3^* \oplus 1. \quad (26)$$

Extending this $SU(3)$ symmetry to $U(3) = SU(3) \times U(1)/\mathbb{Z}_3$ gives an additional $U(1)$ generator, which can be found to coincide with electric charge. This $U(1)_{em}$ symmetry is generated by the number operator for the system, thereby providing an unusually straightforward explanation of charge quantization. Details may be found in [50].

Finally, readers are encouraged to verify that complex conjugation, $i \mapsto -i$, sends particles to anti-particles, $S^u \leftrightarrow S^d$. This parallels our earlier findings for $\mathbb{C} \otimes \mathbb{H}$ where $i \mapsto -i$ similarly gave $\Psi_L \leftrightarrow \Psi_R$.

### 4.5 Minimal left ideals in the matrix formalism

For those more comfortable with the language of matrices, we point out that $S^u$ and $S^d$ may be formulated as

$$S^u + S^d \mapsto \begin{pmatrix} \mathcal{V} & 0 & 0 & 0 & 0 & 0 & 0 & \mathcal{E}^- \\ \bar{\mathcal{D}}^r & 0 & 0 & 0 & 0 & 0 & 0 & \bar{\mathcal{U}}^r \\ \bar{\mathcal{D}}^g & 0 & 0 & 0 & 0 & 0 & 0 & \bar{\mathcal{U}}^g \\ \bar{\mathcal{D}}^b & 0 & 0 & 0 & 0 & 0 & 0 & \bar{\mathcal{U}}^b \\ \mathcal{U}^r & 0 & 0 & 0 & 0 & 0 & 0 & \mathcal{D}^r \\ \mathcal{U}^g & 0 & 0 & 0 & 0 & 0 & 0 & \mathcal{D}^g \\ \mathcal{U}^b & 0 & 0 & 0 & 0 & 0 & 0 & \mathcal{D}^b \\ \mathcal{E}^+ & 0 & 0 & 0 & 0 & 0 & 0 & \bar{\mathcal{V}} \end{pmatrix}. \quad (27)$$

Written in this way, it becomes obvious that there are in fact eight linearly independent minimal left ideals which can be built within $\mathbb{C}l(6)$. Readers interested in finding additional generations of quarks and leptons within $\mathbb{C}l(6)$ should consult [49], Section 9.6 of [46] and [51].

For an interesting proposal connecting this one-generation model to braids, see [52].

### 4.6 Towards weak isospin

As with the example of $\Psi_L$ and $\Psi_R$, the two minimal left ideals $S^u$ and $S^d$ may be transformed into each other under *right* multiplication. However, this time, it is the ladder operators $\omega$ and $\omega^\dagger$ which effect these transitions, generating another copy of $\mathbb{C}l(2)$. Taking the sum, $S_{16} \equiv S^u + S^d$, we then have a faithful representation of $\mathbb{C}l(6) \otimes_{\mathbb{C}} \mathbb{C}l(2)$,

$$
\begin{array}{ccc}
\mathbb{C} \otimes \mathbb{O} & \triangleright \ S_{16} \ \triangleleft & \mathbb{C}l(2)_\omega \\
\downarrow & & \downarrow \\
\mathbb{C}l(6) & & \mathbb{C}l(2) \\
\text{colour, etc} & & \text{isospin type.}
\end{array} \quad (28)
$$

As $\mathbb{C}l(6)$ operators acting on $f \in \mathbb{C} \otimes \mathbb{O}$, schematically, we have $S^u \omega f \sim S^d f$ and $S^d \omega^\dagger f \sim S^u f$.

These two ladder operators, $\omega$ and $\omega^\dagger$, can be seen to induce transitions between isospin pairs, eg $\mathcal{V}$ and $\mathcal{E}^-$, $\mathcal{U}^r$ and $\mathcal{D}^r$, etc. However, this description of weak isospin is clearly not complete in that there is nothing at this stage to indicate that these objects should act on only left-handed states. This will be addressed in the next section.

Before moving on, it should be noted that earlier papers have been found which have much in common with the octonionic model presented here. In 1973, Günaydin and Gürsey found the $SU(3)_C$ structure for a triplet of quarks and anti-quarks using the split octonions [8]. Subsequently in 1977, Barducci et al. [9], built a one-generation model of quarks and leptons from [8], based on $\mathbb{C}l(6)$ and $\mathbb{C}l(2)$. The main differences between the model presented here and [9] lie in the way that particles and anti-particles are related, and in how transitions between isospin states occur. Because of our use of octonionic minimal left ideals, particles and anti-particles are related here simply by $i \mapsto -i$. Furthermore, as we have constructed our quark and lepton states within the space of octonionic maps, not as column vectors, we may then find objects $\omega$ and $\omega^\dagger$ which automatically have the correct electric charges, without having to implement these characteristics by hand. Finally, in the case of $U(1)_{em}$, we find that electric charge is proportional to the number operator for the system, as opposed to being given by a difference between number operators of two distinct Clifford algebras.

### 4.7 $\mathbb{C} \otimes \mathbb{O}$ summary

In this section, we showed that the left action of $\mathbb{C} \otimes \mathbb{O}$ on itself gives a faithful representation of the Clifford algebra $\mathbb{C}l(6)$. We then identified anti-commuting ladder operators generating $\mathbb{C}l(6)$, and used them to build a pair of minimal left ideals, $S^u$ and $S^d$. Under $SU(3)$ ladder symmetry, these ideals were found to transform as do the quarks and leptons of one generation of standard model particles. When this $SU(3)$ symmetry is further extended to $U(3)$, we then find an additional $U(1)$ factor, generated by electric charge [50]. Combining $S^u$ with $S^d$ then provided a faithful representation of $\mathbb{C}l(6) \otimes_{\mathbb{C}} \mathbb{C}l(2)$, where the additional $\mathbb{C}l(2)$ factor enables transitions between isospin up- and down-type states.

## 5 $\mathbb{C}l(4)$: weak isospin

Readers are encouraged to also see a closely related model by Woit [31], which addresses weak isospin in the context of supersymmetric quantum mechanics. For a recent review article on $\mathbb{C}l(4)$ and electroweak theory, see [53].

### 5.1 Clifford algebraic structure

We will now draw the reader's attention to the right action on $\Psi_L + \Psi_R$, and the right action on $S^u + S^d$, which each generated a copy of $\mathbb{C}l(2)$. Recall that the right action on $\Psi_L + \Psi_R$ induced transitions between chiralities $L$ and $R$, while the right action on $S^u + S^d$ induced transitions between isospin up- and down-type states. Together, these two $\mathbb{C}l(2)$ right actions form $\mathbb{C}l(2) \otimes_{\mathbb{C}} \mathbb{C}l(2) \simeq \mathbb{C}l(4)$. Readers should note that this $\mathbb{C}l(4)$ is conceptually distinct from what we have seen before, in that it effects transitions on the space of idempotents.

Nonetheless, we will now work through the same construction with $\mathbb{C}l(4)$ as we did in previous sections. Generators of this $\mathbb{C}l(4)$ may be carefully chosen as

$$\{\tau_1 i\epsilon_1, \tau_2 i\epsilon_1, \tau_3 i\epsilon_1, i\epsilon_2\}, \quad (29)$$

where $\tau_1 \equiv \omega + \omega^\dagger$, $\tau_2 \equiv i\omega - i\omega^\dagger$, $\tau_3 \equiv \omega\omega^\dagger - \omega^\dagger\omega$. It should be noted at this point that the behaviour of our generators (29) under complex conjugation differs from that of previous sections.

As before, these generators may be rewritten in a new basis given by $\{\beta_1, \beta_2, \beta_1^\ddagger, \beta_2^\ddagger\}$ where

$$\beta_1 \equiv \frac{1}{2}\left(-\epsilon_2 + i\epsilon_1\tau_3\right), \quad \beta_2 \equiv \omega^\dagger i\epsilon_1. \quad (30)$$

Here, $\ddagger$ maps $i \mapsto -i$, $\epsilon_j \mapsto -\epsilon_j$ for $j = 1, 2, 3$, and $e_k \mapsto -e_k$ for $k = 1, \ldots 7$, while reversing the order of multiplication. It is then not difficult to confirm that

$$\{\beta_i, \beta_j\} = \{\beta_i^\ddagger, \beta_j^\ddagger\} = 0, \qquad \{\beta_i, \beta_j^\ddagger\} = \delta_{ij}, \quad (31)$$

$\forall\ i = 1, 2$ and $j = 1, 2$.

### 5.2 Leptons as minimal right ideals

With ladder operators defined, we may now construct an algebraic vacuum state as $v_w \equiv \beta_1^\ddagger \beta_2^\ddagger \beta_2 \beta_1$. From $v_w$, we then obtain a minimal right ideal as $\mathcal{L} \equiv v_w \mathbb{C}l(4)$,

$$\mathcal{L} = \mathcal{V}_R v_w + \mathcal{V}_L v_w \beta_1^\ddagger + \mathcal{E}_L^- v_w \beta_2^\ddagger + \mathcal{E}_R^- v_w \beta_1^\ddagger \beta_2^\ddagger, \quad (32)$$

where $\mathcal{V}_R, \mathcal{V}_L, \mathcal{E}_L^-, \mathcal{E}_R^-$ are suggestively named coefficients $\in \mathbb{C}$.

Swapping $\beta_i \leftrightarrow \beta_i^\ddagger$ and defining $v'_w \equiv \beta_1 \beta_2 \beta_2^\ddagger \beta_1^\ddagger$ gives a linearly independent minimal right ideal as

$$\gamma^0 \bar{\mathcal{L}} \equiv \left( \bar{\mathcal{V}}_L v'_w - \bar{\mathcal{V}}_R v'_w \beta_1 + \mathcal{E}_R^+ v'_w \beta_2 - \mathcal{E}_L^+ v'_w \beta_1 \beta_2 \right) i\epsilon_1. \quad (33)$$

The bars over top of the variables here, as in $\bar{\mathcal{L}}$ and $\bar{\mathcal{V}}_L$, are meant only to identify anti-particles; they do not imply hermitian conjugation and multiplication by $\gamma^0$.

It should be noted that in this particular $\mathbb{C}l(4)$ construction (29), swapping $\beta_i \leftrightarrow \beta_i^\ddagger$ provided a new minimal right ideal, $\gamma^0 \bar{\mathcal{L}}$, which automatically includes a factor of $\gamma^0$. This additional $\gamma^0$ will be familiar from QFT kinetic terms of the form $\psi^\dagger \gamma^0 \gamma^\mu \partial_\mu \psi$.

### 5.3 $SU(2)$ ladder symmetry

In parallel with the previous section, $SU(2)$ symmetries may now be applied to our ladder operators as

$$e^{-ir_j T_j} \beta_k e^{ir_j T_j} \quad \text{and} \quad e^{-ir_j T_j} \beta_k^\ddagger e^{ir_j T_j}. \quad (34)$$

The three $T_j$ generate $SU(2)$, and are found to be

$$T_1 \equiv \tau_1 \tfrac{1}{2}(1+i\epsilon_3), \quad T_2 \equiv \tau_2 \tfrac{1}{2}(1+i\epsilon_3), \quad T_3 \equiv \tau_3 \tfrac{1}{2}(1+i\epsilon_3). \quad (35)$$

As before, transformations on the ladder operators induce transformations on our minimal right ideals. So we then find that under $SU(2)$, the ideals $\mathcal{L}$ and $\gamma^0 \bar{\mathcal{L}}$ transform as

$$\mathcal{L} \sim 1 \oplus 2 \oplus 1, \quad \gamma^0 \bar{\mathcal{L}} \sim 1 \oplus 2 \oplus 1. \quad (36)$$

These transformation properties agree with the leptonic identifications we made in Eqs. (32) and (33), which will be further justified in the next section.

So here we have again found standard model group representations by taking the special unitary symmetries of our ladder operators, and examining the transformations they induce on minimal one-sided ideals. In this case, we have found the behaviour of leptons under $SU(2)_L$. It is worth emphasizing here that when the $SU(2)$ symmetry of these ladder operators was applied to minimal right ideals, it *acted automatically on states of only a single chirality*. We did not need to implement a projector by hand. Please see Fig. 4.

Finally, we point out that if a $U(2)$ ladder symmetry is taken in place of $SU(2)$, then we find an extra $U(1)$ charge, given that $U(2) = SU(2) \times U(1)/\mathbb{Z}_2$. This $U(1)$ is again generated by the system's number operator, and in this case, coincides with weak hypercharge [46,53].

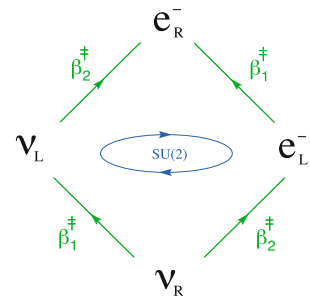

**Fig. 4** The leptonic minimal right ideal $\mathcal{L}$. As with our previous examples, this ideal resembles a Fock space, with the right-handed neutrino acting as the (formal) vacuum state. It should be noted that the $SU(2)$ symmetries of our ladder operators are found to act automatically on lepton states of only a single chirality. That is, without the need to impose a chiral projector by hand

### 5.4 $\mathbb{C}l(4)$ summary

In this section, we focussed in on a representation of $\mathbb{C}l(4)$ which induces transitions of isospin and chirality idempotents. We identified a particular set of anti-commuting ladder operators generating $\mathbb{C}l(4)$, and used it to build a pair of minimal right ideals, $\mathcal{L}$ and $\gamma^0 \bar{\mathcal{L}}$. Under the $SU(2)$ symmetry of these ladder operators, the ideals were found to transform as do the leptons of one generation of standard model particles, together with a right-handed neutrino. Here, the group $SU(2)$ was found to act automatically on states of only a single chirality. Finally, when a $U(2)$ ladder symmetry was used in place of $SU(2)$, we then found an additional $U(1)$ generator with eigenvalues consistent with weak hypercharge.

## 6 All together: ladder symmetries to $SU(3)_C \times SU(2)_L \times U(1)_Y / \mathbb{Z}_6$

We have just come from finding familiar standard model particle representations by considering the action of division algebras on themselves. That is, the special unitary symmetries of division algebraic ladder operators led to $SU(3)_C$ and $SU(2)_L$ when acting on minimal one-sided ideals. Our goal now is to combine these fragments into a single model.

### 6.1 Clifford algebraic structure

It is well known that the Clifford algebra $\mathbb{C}l(10)$ can provide the background structure for $Spin(10)$ and $SU(5)$ grand unified theories [4]. We will then use our division algebraic actions from previous sections to build a representation of $\mathbb{C}l(6) \otimes_\mathbb{C} \mathbb{C}l(4) \simeq \mathbb{C}l(10)$.

Making use of equations (19) and (30), let us define ten ladder operators to be

$$\begin{aligned} &A_1 \equiv a_1|\mathbb{I}, \quad A_2 \equiv a_2|\mathbb{I}, \quad A_3 \equiv a_3|\mathbb{I}, \\ &B_1 \equiv ie_7|\beta_1, \quad B_2 \equiv ie_7|\beta_2, \\ &A_1^\ddagger \equiv a_1^\dagger|\mathbb{I}, \quad A_2^\ddagger \equiv a_2^\dagger|\mathbb{I}, \quad A_3^\ddagger \equiv a_3^\dagger|\mathbb{I}, \\ &B_1^\ddagger \equiv ie_7|\beta_1^\ddagger, \quad B_2^\ddagger \equiv ie_7|\beta_2^\ddagger, \end{aligned} \quad (37)$$

where $\mathbb{I}$ represents the identity. Given Eqs. (21) and (31), it is trivial to confirm that these obey the usual anti-commutation relations.

Readers should note that from the perspective of the Clifford algebra alone, there is no real distinction between the $A_i$ operators and the $B_j$ operators. However the same cannot be said when the $A_i$ and $B_j$ operators are realised in terms of division algebras, as they were in this article. That is, the $B_j$ may be considered as truly distinct from the $A_i$, in that the $B_j$ were introduced so as to effect transitions between idempotents. Said another way, the $A_i$ can be seen to map a left ideal to itself, whereas the $B_j$ were introduced so as to map one ideal to another.

### 6.2 One generation as minimal ideals

Using the same procedure as before, we may now construct a vacuum state as

$$v_t \equiv A_1 A_2 A_3 B_1 B_2 B_2^\ddagger B_1^\ddagger A_3^\ddagger A_2^\ddagger A_1^\ddagger = v_c | v_w, \quad (38)$$

from which we build our minimal left ideal,

$$\begin{aligned} S &\equiv \mathbb{C}l(10) v_t \\ &= \mathcal{V}_R v_t + \bar{\mathcal{D}}_L^i A_i^\ddagger v_t + \mathcal{V}_L B_1^\ddagger v_t + \mathcal{E}_L^- B_2^\ddagger v_t \\ &\quad + \mathcal{U}_R^k \epsilon_{ijk} A_j^\ddagger A_i^\ddagger v_t + \bar{\mathcal{D}}_R^i A_i^\ddagger B_1^\ddagger v_t \\ &\quad + \bar{\mathcal{U}}_R^i A_i^\ddagger B_2^\ddagger v_t + \mathcal{E}_R^- B_2^\ddagger B_1^\ddagger v_t \\ &\quad + \mathcal{E}_L^+ A_3^\ddagger A_2^\ddagger A_1^\ddagger v_t + \mathcal{U}_L^k \epsilon_{ijk} B_1^\ddagger A_j^\ddagger A_i^\ddagger v_t \\ &\quad + \mathcal{D}_L^k \epsilon_{ijk} B_2^\ddagger A_j^\ddagger A_i^\ddagger v_t + \bar{\mathcal{U}}_L^i B_2^\ddagger B_1^\ddagger A_i^\ddagger v_t \\ &\quad + \mathcal{E}_R^+ A_3^\ddagger A_2^\ddagger A_1^\ddagger B_1^\ddagger v_t + \bar{\mathcal{V}}_R A_3^\ddagger A_2^\ddagger A_1^\ddagger B_2^\ddagger v_t \\ &\quad + \mathcal{D}_R^k \epsilon_{ijk} B_2^\ddagger B_1^\ddagger A_j^\ddagger A_i^\ddagger v_t \\ &\quad + \bar{\mathcal{V}}_L B_2^\ddagger B_1^\ddagger A_3^\ddagger A_2^\ddagger A_1^\ddagger v_t. \end{aligned} \quad (39)$$

As with $\gamma^0 \bar{\mathcal{L}}$ of Eq. (33), the antiparticles within $S$ can be seen to include a factor of $\gamma^0$ automatically. The complex coefficients, $\mathcal{V}_R, \bar{\mathcal{D}}_L^i, \ldots$, are written here so as to anticipate how these states will eventually transform under $SU(3)_C \times SU(2)_L \times U(1)_Y/\mathbb{Z}_6 \subset SU(5)$.

As before, this minimal left ideal exhibits the structure of a Fock space. Readers may also notice that removing the idempotent $v_t$ from these states leaves us with the exterior algebra $\Lambda \mathbb{C}^5$, as described in [4].

### 6.3 From $SU(5)$ ladder symmetry to $SU(3)_C \times SU(2)_L \times U(1)_Y/\mathbb{Z}_6$

It is straightforward to see that the special unitary transformations on $\mathbb{C}l(10)$ ladder operators give $SU(5)$. These $SU(5)$ ladder symmetries further induce transformations on the minimal left ideal (39). Readers are encouraged to confirm that the action of $SU(5)$ on (39) coincides exactly with the $SU(5)$ transformations of particles and anti-particles in the Georgi–Glashow model.

In total, there are 24 generators of $SU(5)$ ladder symmetries, which split into two types. The first type of generator mixes $A$- and $B$-type ladder operators. These will be known as *mixing* generators, and can be written using the hermitian forms

$$A_j^\ddagger B_k + B_k^\ddagger A_j \quad \text{and} \quad i A_j^\ddagger B_k - i B_k^\ddagger A_j. \quad (40)$$

Since $j$ runs from 1 to 3 and $k$ runs from 1 to 2, we find 12 generators of this first type.

Now, as far as the Clifford algebra $\mathbb{C}l(10)$ is concerned, we have no reason to exclude these generators from consideration. However, from the perspective of our division algebraic construction, $A$- and $B$-type ladder operators are clearly algebraically distinct. We will then exclude these 12 elements from this model. Incidentally, it is precisely this first type of generator which is responsible for proton decay.

The second type of generator does not mix $A$- and $B$-type ladder operators. In total, there are 12 such generators remaining. The first eight are given by

$$\begin{aligned} \Lambda_1 | \mathbb{I} &= -A_2^\ddagger A_1 - A_1^\ddagger A_2, \\ \Lambda_2 | \mathbb{I} &= i A_2^\ddagger A_1 - i A_1^\ddagger A_2, \\ \Lambda_3 | \mathbb{I} &= A_2^\ddagger A_2 - A_1^\ddagger A_1, \\ \Lambda_4 | \mathbb{I} &= -A_1^\ddagger A_3 - A_3^\ddagger A_1, \\ \Lambda_5 | \mathbb{I} &= -i A_1^\ddagger A_3 + i A_3^\ddagger A_1, \\ \Lambda_6 | \mathbb{I} &= -A_3^\ddagger A_2 - A_2^\ddagger A_3, \\ \Lambda_7 | \mathbb{I} &= i A_3^\ddagger A_2 - i A_2^\ddagger A_3, \\ \Lambda_8 | \mathbb{I} &= -\frac{1}{\sqrt{3}} \left( A_1^\ddagger A_1 + A_2^\ddagger A_2 - 2 A_3^\ddagger A_3 \right), \end{aligned} \quad (41)$$

which can be seen to generate $SU(3)_C$ when applied to the minimal left ideal (39). Here, the $\Lambda_j$ are defined as in equation (25). The next three generators are given by

$$\begin{aligned} \mathbb{I} | T_1 &= B_1^\ddagger B_2 + B_2^\ddagger B_1, \\ \mathbb{I} | T_2 &= i B_2^\ddagger B_1 - i B_1^\ddagger B_2, \\ \mathbb{I} | T_3 &= B_1^\ddagger B_1 - B_2^\ddagger B_2, \end{aligned} \quad (42)$$

which generate $SU(2)_L$ when applied to (39). Here, the $T_j$ are defined as in equation (35). Finally, the twelfth generator is realised as

$$\frac{1}{3}\left(\sum_j A_j^\ddagger A_j\right) - \frac{1}{2}\left(\sum_k B_k^\ddagger B_k\right), \tag{43}$$

and can be seen to assign charges to (39) which coincide with hypercharge, $Y$.

*Hence, we have found that it is exactly the non-mixing SU(5) ladder symmetries which generate the standard model's gauge group, $SU(3)_C \times SU(2)_L \times U(1)_Y / \mathbb{Z}_6$.*

Finally, as with previous cases, we may consider $U(5) = SU(5) \times U(1)/\mathbb{Z}_5$ ladder symmetries, as opposed to $SU(5)$. This again introduces an extra $U(1)$ generator, proportional to the number operator. Up to an overall phase, this number operator,

$$N_5 \equiv \sum_j A_j^\ddagger A_j + \sum_k B_k^\ddagger B_k, \tag{44}$$

is found to give the $X$ charges from the well-known symmetry breaking pattern $Spin(10) \mapsto SU(5) \times U(1)_X/\mathbb{Z}_5$ of [54]. Explicitly,

$$X = 2N_5 - 5 = 5(B-L) - 4Y, \tag{45}$$

where we are taking $Y$ according to the weak hypercharge conventions of [5].

### 6.4 Summary

In this section, we combined our previous $\mathbb{C}l(4)$ and $\mathbb{C}l(6)$ results so as to yield a division algebraic representation of $\mathbb{C}l(10)$. From $\mathbb{C}l(10)$ ladder operators, we constructed a 32-$\mathbb{C}$-dimensional minimal left ideal. Then, under special unitary ladder symmetries, this minimal left ideal was found to transform as do the particles and anti-particles of the $SU(5)$ Georgi–Glashow grand unified theory.

Finally, under the requirement that $A$- and $B$-type division algebraic actions be kept distinct, we find that the $SU(5)$ ladder symmetries then reduce immediately to $SU(3)_C \times SU(2)_L \times U(1)_Y/\mathbb{Z}_6$.

Making use of the full $U(5)$ symmetry leads us to the same result, but introduces the possibility of an extra (presumably gauged) $U(1)_X$.

## 7 Outlook

We have come from demonstrating how four low-dimensional algebras: $\mathbb{R}$ (1D), $\mathbb{C}$ (2D), $\mathbb{H}$ (4D), and $\mathbb{O}$ (8D), can act on themselves so as to yield group representations of the Georgi–Glashow model. Here, the group $SU(5)$ arises as symmetries of Clifford algebraic ladder operators. We point out, though, that only half of these $SU(5)$ generators preserve the underlying algebraic structure. Perhaps unexpectedly, we find that it is precisely this subset which generates the standard model gauge group, $G_{sm} = SU(3)_C \times SU(2)_L \times U(1)_Y/\mathbb{Z}_6$.

It bears mentioning that the reduction of $SU(5) \mapsto SU(3)_C \times SU(2)_L \times U(1)_Y/\mathbb{Z}_6$ has not been mediated here by a Higgs boson. Instead, we emphasize that the full $SU(5)$ symmetry should never be fully realised in this division algebraic model in the first place.

Finally, we point out one last avenue worth investigation. Readers may have noticed that with the introduction of $\omega \equiv a_1 a_2 a_3$, and $\omega^\dagger = a_3^\dagger a_2^\dagger a_1^\dagger$, we were able to show that sequences of complex octonions can behave as $\mathbb{C}l(2) \simeq \mathbb{C} \otimes \mathbb{H}$. In other words, new algebraic behaviour can arise at different chain lengths of the original algebra (length three in this case).

This algebraic phenomenon bears resemblance to the emergence of effective theories in physics at different energy scales. We might then ask if collective algebraic behaviour might ultimately be used to address currently unexplained physical phenomena, such as colour confinement.

**Acknowledgements** *In remembrance of October 9, 2012, the day that silenced one fiery soul, and amplified another*. A special thank you to Ben Allanach, Latham Boyle, John Huerta, Judd Harrison, Ciaran Hughes, Mia Hughes, Carlos Tamarit, Paul Townsend, and friends at AQG 2016 for helpful feedback.
This work has been partially supported by STFC consolidated Grant ST/P000681/1. The author is furthermore grateful for support from the NSERC Postdoctoral Fellowship, and from the Walter Grant Scott Research Fellowship in Physics at Trinity Hall, University of Cambridge.

## References

1. H. Georgi, S. Glashow, Unity of all elementary-particle forces. Phys. Rev. Lett. **32**(8), 438–441 (1974)
2. K. Sundermeyer, *Symmetries in fundamental physics* (Springer, New York, 2014)
3. E. Witten, Grand unification with and without supersymmetry, in *Introduction to Supersymmetry in Particle and Nuclear Physics*, ed. by O. Castaos, A. Frank, L. Urrutia (Springer, Boston, 1984), pp. 53–76
4. J. Baez, J. Huerta, The algebra of grand unified theories. Bull. Am. Math. Soc. **47**, 483–552 (2010). arXiv:0904.1556 [hep-th]
5. P. Langacker, *The standard model and beyond* (CRC Press, Boca Raton, 2010)

6. The Super-Kamiokande Collaboration, Search for proton decay via $p \to \nu K^+$ using 260 kiloton-year data of Super-Kamiokande. Phys. Rev. D **90**, 072005 (2014)
7. The Super-Kamiokande Collaboration, Search for proton decay via $p \to e^+\pi^0$ and $p \to \mu^+\pi^0$ in 0.31 megaton·years exposure of the Super-Kamiokande water Cherenkov detector (2016). arXiv:1610.03597 [hep-ex]
8. M. Günaydin, F. Gürsey, Quark structure and the octonions. J. Math. Phys. **14**(11), 1651–1667 (1973)
9. A. Barducci, F. Buccella, R. Casalbuoni, L. Lusanna, E. Sorace, Quantized Grassmann variables and unified theories. Phys. Lett. **67B**, 344 (1977)
10. R. Casalbuoni, R. Gatto, Unified theories for quarks and leptons based on Clifford algebras. Phys. Lett. B **90**, 81–86 (1979)
11. G. Dixon, *Division algebras: octonions, quaternions, complex numbers and the algebraic design of physics* (Kluwer Academic Publishers, Amsterdam, 1994)
12. S.L. Adler, Coleman-Weinberg symmetry breaking in SU(8) induced by a third rank antisymmetric tensor scalar field. J. Phys. A. **49**(31), 315401 (2016). arXiv:1602.05212 [hep-ph]
13. H. Sonmez, The flipped SU(5) string vacua classification: a variation of the SO(10) breaking basis vector. Phys. Rev. D **93**, 125002 (2016). arXiv:1603.03504 [hep-th]
14. F.F. Deppisch, C. Hati, S. Patra, U. Sarkar, J.W.F. Valle, 331 models and grand unification: from minimal SU(5) to minimal SU(6). Phys. Lett. B **762**, 432–440 (2016). arXiv:1608.05334 [hep-ph]
15. K. Kojima, K. Takenaga, T. Yamashita, Gauge symmetry breaking patterns in an SU(5) grand gauge-Higgs unification. Phys. Rev. D **95**, 015021 (2017). arXiv:1608.05496 [hep-ph]
16. C.H. Lee, R.N. Mohapatra, Vector-like quarks and leptons, $SU(5) \times SU(5)$ grand unification, and proton decay. JHEP **02**, 080 (2017). arXiv:1611.05478 [hep-ph]
17. N. Maekawa, Y. Muramatsu, Flavor changing nucleon decay. Phys. Lett. B. **767**, 398–402 (2017). arXiv:1601.04789 [hep-ph]
18. A. Athenodorou, M. Teper, On the mass of the world-sheet 'axion' in SU(N) gauge theories in 3+1 dimensions. Phys. Lett. B **771**, 408–414 (2017). arXiv:1702.03717 [hep-lat]
19. G.K. Karananas, M. Shaposhnikov, Gauge coupling unification without leptoquarks. Phys. Lett. B **771**, 332–338 (2017). arXiv:1703.02964 [hep-ph]
20. L. Randall, D. Simmons-Duffin, Quark and lepton flavor physics from F-theory. arXiv:0904.1584 [hep-ph]
21. J.J. Heckman, Particle physics implications of F-theory. Ann. Rev. Nucl. Part. Sci. **60**, 237–265 (2010). arXiv:1001.0577 [hep-th]
22. T. Abe, Y. Fujimoto, T. Kobayashi, T. Miura, K. Nishiwaki, M. Sakamoto, Y. Tatsuta, Classification of three-generation models on magnetized orbifolds. Nucl. Phys. B **894**, 374–406 (2015). arXiv:1501.02787 [hep-ph]
23. B. Assel, S. Schafer-Nameki, Six-dimensional origin of $\mathcal{N} = 4$ SYM with duality defects. JHEP **12**, 058 (2016). arXiv:1610.03663 [hep-th]
24. K. Krasnov, Fermions, differential forms and doubled geometry. arXiv:1803.06160 [hep-th]
25. A.H. Chamseddine, A. Connes, W.D. van Suijlekom, Beyond the spectral standard model: emergence of Pati-Salam unification. JHEP **1311**, 132 (2013). arXiv:1304.8050 [hep-th]
26. A. Devastato, F. Lizzi, C. Valcarcel Flores, D. Vassilevich, Unification of coupling constants, dimension six operators and the spectral action. Int. J. Mod. Phys. A **30**, 1550033 (2015). arXiv:1410.6624 [hep-ph]
27. U. Aydemir, D. Minic, C. Sun, T. Takeuchi, Pati-Salam unification from non-commutative geometry and the TeV-scale $W_R$ boson. Int. J. Mod. Phys. A **31**, 1550223 (2016). arXiv:1509.01606 [hep-ph]
28. N. Bizi, C. Brouder, F. Besnard, Space and time dimensions of algebras with applications to Lorentzian noncommutative geometry and the standard model. arXiv:1611.07062 [hep-th]
29. L. Dabrowski, F. D'Andrea, A. Sitarz, The standard model in non-commutative geometry: fundamental fermions as internal forms. arXiv:1703.05279 [math-ph]
30. T. Kugo, P. Townsend, Supersymmetry and the division algebras. Nucl. Phys. B **221**, 357–380 (1983)
31. P. Woit, Supersymmetric quantum mechanics, spinors, and the standard model. Nucl. Phys. B **303**, 329–342 (1988)
32. J.M. Evans, Supersymmetric (Yang–Mills) theories and division algebras. Nucl. Phys. B **298**, 92–108 (1988)
33. B. Bajc, J. Hisano, T. Kuwahara, Y. Omura, Threshold corrections to dimension-six proton decay operators in non-minimal SUSY SU(5) GUTs. Nucl. Phys. B **910**, 1–22 (2016). arXiv:1603.03568 [hep-ph]
34. J. Ellis, J.L. Evans, N. Nagata, D.V. Nanopoulos, K.A. Olive, No-scale SU(5) super-GUTs. Eur. Phys. J. C **77**(4), 232 (2017). arXiv:1702.00379 [hep-ph]
35. M.J. Duff, S. Ferrara, A. Marrani, $H = 3$ unification of curious supergravities. JHEP **01**, 023 (2017). arXiv:1610.08800 [hep-th]
36. T. Kobayashi, Y. Omura, O. Seto, K. Ueda, Realization of a spontaneous gauge and supersymmetry breaking vacuum (2017). arXiv:1705.00809 [hep-ph]
37. T. Dray, C.A. Manogue, Using octonions to describe fundamental particles. Prog. Math. Phys. **34**, 451–466 (2004)
38. K. Morita, Algebraic gauge theory of quarks and leptons. JPS Conf. Proc. **7**, 010010 (2015)
39. M. Dubois-Violette, Exceptional quantum geometry and particle physics. Nucl. Phys. B. **912**, 426–449 (2016). arXiv:1604.01247 [math.QA]
40. C. Burdik, S. Catto, Y. Gurcan, A. Khalfan, L. Kurt, Revisiting the role of octonions in hadronic physics. Phys. Part. Nucl. Lett. **14**(2), 390–394 (2017)
41. L. Basso, Phenomenology of the minimal $B - L$ extension of the standard model at the LHC, Ph.D. thesis, University of Southampton. arXiv:1106.4462 [hep-ph]
42. J. Heeck, Unbroken $B - L$ symmetry. Phys. Lett. B **739**, 256–262 (2014). arXiv:1408.6845
43. L. Boyle, S. Farnsworth, Rethinking Connes' approach to the standard model of particle physics via non-commutative geometry. N. J. Phys. 17, 023021 (2015). arXiv:1408.5367 [hep-th]
44. P. Pérez, C. Murgui, Sterile neutrinos and $B - L$ symmetry. arXiv:1708.02247 [hep-ph]
45. R. Abłamowicz, *Construction of spinors via Witt decomposition and primitive idempotents: a review, Clifford Algebras and Spinor Structures* (Kluwer Academic Publishers, Dordrecht, 1995)
46. C. Furey, Standard model physics from an algebra? Ph.D. thesis, University of Waterloo (2015). https://www.repository.cam.ac.uk/handle/1810/254719. arXiv:1611.09182 [hep-th]
47. S. De Leo, Quaternions for GUTs. Int. J. Theor. Phys. **35**, 1821 (1996)
48. P. Bolokhov, Quaternionic wavefunction. arXiv:1712.04795
49. C. Furey, Generations: three prints, in colour. JHEP **10**, 046 (2014). arXiv:1405.4601 [hep-th]
50. C. Furey, Charge quantization from a number operator. Phys. Lett. B **742**, 195–199 (2015). arXiv:1603.04078 [hep-th]
51. C. Furey, Three generations, two unbroken gauge symmetries, and one eight-dimensional algebra (2018) (**in preparation**)
52. N. Gresnigt, Braids, normed division algebras, and standard model symmetries. arXiv:1803.02202
53. C. Furey, A demonstration that electroweak theory can violate parity automatically (leptonic case). Int. J. Mod. Phys. A. **33**(04), 1830005 (2018)
54. R. Slansky, *Group theory for unified model building, Physics reports*, vol. 79 (North-Holland Publishing Company, New York, 1981)